\documentstyle[aps,prl,twocolumn,epsfig]{revtex}
\begin{document}
\title{
Order of the phase transition in models of DNA thermal denaturation}
\author{ Nikos Theodorakopoulos$^{1,2}$, Thierry Dauxois$^{2}$
and Michel Peyrard$^{2}$ }
\address{$^{1}$ Theoretical and Physical Chemistry Institute,
National Hellenic Research Foundation\\
Vasileos Constantinou 48, 116 35 Athens, Greece\\
 $^{2}$ Laboratoire de Physique, UMR-CNRS 5672, ENS Lyon,
46 All\'{e}e d'Italie, 69364 Lyon C\'{e}dex 07, France }
\date{submitted to Physical Review Letters on January 5, 2000; 
revised March 31, 2000}
\draft
\maketitle
\begin{abstract}
We examine the behavior of a model which describes the
melting of double-stranded DNA chains. The model, 
with displacement-dependent stiffness constants and a Morse
on-site potential, is analyzed numerically; depending on the 
stiffness parameter, it is shown to to have either 
(i) a second order transition with $\nu_{\perp}=-\beta=1, 
\nu_{||}=\gamma/2=2$ (characteristic of short-range attractive
part of the Morse potential) or,
(ii) a first-order transition with finite melting entropy,
discontinuous fraction of bound pairs, divergent correlation lengths,
and critical exponents $\nu_{\perp}=-\beta=1/2, \nu_{||}=\gamma/2=1$.
\end{abstract}
Early models of DNA thermal denaturation, i.e. the separation of the
two strands upon heating\cite{WARTELLBENIGHT},
were based on an Ising-like
description in which a base-pair was either closed or open;
the relative tendencies of the system to occupy one of the two
states were introduced explicitly, in terms of
temperature-dependent free enthalpies.  As a consequence,
although a judicious choice of such enthalpies has proved
useful in describing some aspects of DNA denaturation\cite{PoSche},
understanding of this remarkable one-dimensional cooperative
phenomenon in terms of standard statistical mechanics -i.e.
a Hamiltonian model with temperature-independent parameters-
remained an outstanding problem.\par
More recent research has emphasized the role
of the large amplitude fluctuations that precede the transition and the
intrinsically nonlinear mechanisms which are needed to
describe such fluctuations\cite {Englander}.
In such models the status of a base pair is characterized by the
distance
between the two bases. An on-site asymmetric potential with a flat part
at large values of the displacement emulates the tendency of
the pair to ``melt'' at high temperatures, as thermal phonons drive the
particles outside the well and towards the flat portion of the
potential.  \par
In the original version ("type I" model) coupling between successive
base pairs is harmonic\cite{PB};
the resulting path to the melting instability appears smooth; this is at
variance with  the sharp features of the transition observed
experimentally.
A generalization \cite{dauxpeyr1,dauxpeyr2}  of the model to include
displacement-dependent stiffness constants, i.e. "stacking parameters"
which describe the coupling between successive
base pairs, has revealed a dramatic sharpening of the transition.
The predictions of the latter, ("type II") model have
been compared successfully with experimental results\cite{giansanti}.
Furthermore,
investigations of heterogeneous DNA have shown that the model 
yields features of multistep melting similar to those observed in
experiments\cite{CuleHwa}. In fact, the authors of 
Ref. \cite{CuleHwa} have, in passing, pointed out the formal 
analogies between the melting instability of homogeneous DNA 
and other continuous phase transitions, e.g. wetting of a
one-dimensional interface from a substrate, adsorption of polymers
by a solid surface etc.;
in addition, they have demonstrated that "type II" models generate an
"entropic barrier" which is largely responsible for the narrowing of the
transition. Their analysis of the order parameter
led them however to suggest that, in spite of the dramatic
narrowing of the transition, the critical exponent remains unchanged.
As a result, the exact character of the homogeneous
DNA transition remained somewhat elusive.\par
In this Letter we report on the scaling behavior of type-II
model of the denaturation of ideal, homogeneous DNA.
We show that, for values of the stacking parameter used in
\cite{dauxpeyr2} the type-II model exhibits a peculiar type
of first order transition, with a finite melting entropy, 
a discontinuity in the fraction of bound pairs (the usual
DNA observable), and  divergent longitudinal and transverse correlation
lengths \cite{wetting}; as the value of the stacking parameter changes,
and the range of the "entropic barrier" becomes shorter than,
or comparable to the range of the Morse potential, the
transition changes to second order, as in type-I models. \par
The Hamiltonian of the model is
\begin{eqnarray}
H=\sum_{n}^{}\Biggl[ {p_n^2 \over 2 m}
+ W(y_n,y_{n-1}) + V(y_n)  + Dha y_n   \Biggr]
\label{eq:eHamiltonian}
\end{eqnarray}
where $m$ is the reduced mass of a base pair,
$y_n$ denotes the stretching of the hydrogen bonds connecting
the two bases of the n$^{\text{th}}$ pair and $p_n = m (dy_n/dt)$.
Coupling between successive base pairs is
described by $ W(y_n,y_{n-1})={K\over 2} [1+ \rho
e^{-\alpha(y_n+y_{n-1})} ] \ (y_n-y_{n-1})^{2} $. The parameter $\rho$
can take non-zero values in type-II models;
The choice of this coupling potential is motivated by the observation
that the stacking energy is a property of base {\em pairs} rather than
individual bases. The effective coupling constant decreases
from $ (1+\rho)K $ to $K$ when either one of the two interacting base
pairs is open, in qualitative
agreement with the known properties of base-pair interactions in DNA.
The third term in (\ref{eq:eHamiltonian}) stands for an on-site
potential which describes the interaction of the two bases in a pair;
the Morse potential $V(y_n) = D(e^{-ay_n}-1)^2 $ has been chosen
because it has the correct qualitative shape.
Finally, the fourth term, which describes the effect of a transverse,
external stress $h$, is in fact a mathematical device useful in
practical calculations. By letting the dimensionless $h$ approach 
zero from above, it is possible to extract the scaling behavior near 
the transition; at the same time, since the partition function is now
divergence-free at any $h>0$, a source of criticism of the model on
formal mathematical grounds\cite{chinese} is removed.
The parameters of the model are:
$D=0.03$ eV, $K=0.06$ eV/$\AA^2$, $a=4.5\ \AA^{-1}$,
$\alpha=0.35\ \AA^{-1} $, $m=300$ a.m.u.
and the lattice constant $l=  4.5\AA$.\par
The thermodynamic properties of (\ref{eq:eHamiltonian})
can be described  \cite {KRUMSCHRIEF} in terms of  the eigenvalues
and eigenstates of the transfer-integral (TI) equation
\begin{equation} a \int_{-\infty}^{\infty} d y  e^{ - {\cal K}
(x,y)/k_{B}T }\
\phi_i(y)= e^{- \varepsilon_i  /k_{B}T }\
\phi_i(x) \quad
\label{eq:transferoperator}
\end{equation}
with a symmetrized kernel ${\cal K}(x,y)=  W(x,y) +{1\over 2}
[V(y)+V(x)]+{Dah\over 2} (y+x )    $; here
$T$ is the temperature and $k_{B}$ the Boltzmann constant.
Details of the numerical procedure used in solving
(\ref{eq:transferoperator})
have been given in \cite{dauxpeyr2}; in the present study a
Gauss-Legendre quadrature formula has been used.
In the gradient expansion (continuum) approximation, valid in the
temperature window $D< k_{B}T < K/a^{2}$ \cite{GuyerMiller},
the integral equation (\ref{eq:transferoperator})
can be well approximated by the second-order differential equation
\begin{equation}
-\frac{1}{\delta \> g(x) a^{2} } \frac{d^{2}\phi_{i}}{dx^{2}} +  U(x)
\phi_{i}
 =  {\tilde \epsilon_{i} }\phi_{i}
\label{anhPS}
\end{equation}
where
$\delta\equiv \sqrt{2KD/a^{2}}/(k_B T)$,
$g(x) = 1 + \rho \exp (-2\alpha x)$,
$  D{\tilde \epsilon}_{\i}=  \epsilon_{i}
 + (k_{B}T/2)\ln(2\pi a^{2}k_{B}T /K)$,
$DU(x)=  V(x) + (k_{B}T/2)   \ln g(x) $,
and the limit $h\to 0$ has been  explicitly taken.\par
Of particular interest are (i) the lowest eigenvalue $\epsilon_{0}$,
which, in the thermodynamic limit, is equal to the
free energy per site $f$, (ii) the ground state $\phi_0$, which
determines the order parameter
$\sigma\equiv\langle y \rangle\equiv\int dy y |\phi_{0}| ^{2}$ and its fluctuations
$\langle(\delta y)^{2}\rangle\equiv\int dy (y-\sigma)^{2} |\phi_{0}| ^{2}$
($\equiv \xi_{\perp}^{2}$, the
transverse correlation length in the language of wetting \cite{KroLip}),
and (iii) the next-to-lowest eigenvalue, which controls the longitudinal
correlation length $\xi_{||} =l k_{B}T/ (\epsilon_{1}- \epsilon_{0})$.
Computation of the static structure factor
$ S(q,T)=  \sum_{n}\exp(-iqnl) \langle\delta y_{n}\delta y_{0}\rangle$,
where $\delta y_{n}= y_{n}-\sigma$, requires knowledge of the full
spectrum;
in terms of the matrix elements $M_{i} \equiv \langle i|x|0\rangle$,
and the differences $\Delta_{i} \equiv ( \epsilon_{i}-\epsilon_{0}
)/k_{B}T$,
\begin{equation}
S(q,T) =\sum_{i}{'} |M_{i}|^{2}  \frac { \sinh{\Delta_{i} }     }
{ \cosh{\Delta_{i}  } - \cos(ql)  } ,
\label{sq}
\end{equation}
where the ground state is excluded from the summation.\par
For $\rho=0$ (type - I case)
the asymptotic properties of the denaturation instability
can be obtained from the solution of
the "pseudoSchr\" odinger" Eq.(\ref{anhPS}) with $g=1$ 
\cite{LL,Simon}. In brief: 
as  long as $1> \delta>\delta_{c}=1/2$, there is a
single bound state with
${\tilde \epsilon}_{0}/D=1-(1-\delta_{c}/\delta)^{2}$
which disappears 
 at $\delta=\delta_{c}$, corresponding to a critical temperature
$k_{B}T_{c}=2\sqrt{2KD}/a$;
in the critical regime preceding the
instability, $ |t|\ll1,  t \equiv T/T_{c}-1$, the power laws
 $l/\xi_{||} \propto |t|^{\nu_{||} }$,  $\sigma \propto
|t|^{\beta} $ and    $\xi_{\perp} \propto |t|^{-\nu_{\perp}} $
hold, with $\nu_{||}=2$ and $\nu_{\perp}= - \beta=1$.
Furthermore, we have calculated the 
structure factor\cite{details} in the regime
$qa\ll 1$, $\xi_{||}/l \gg 1$,
\begin{equation}
S(q,T)=  \frac{1}{a^{2}(\delta-\delta_{c})^{2}} \frac{\xi_{||}}{l}
F(q\xi_{||})
\; ,
\end{equation}
where
$F(x)=(1/4)x^{-2}[1-1/\cosh( \text{arcsinh}(x)/2)]$. 
This implies (i) a zero-field isothermal susceptibility 
$\chi = \lim_{h\to 0} a(\partial \sigma / \partial h)_{T}  \propto
|t|^{-\gamma}$,
where $\gamma=4$, and (ii) critical correlations
($q\xi_{||}\gg1$), $S(q) \propto (qla)^{-2+\eta}$ with
$\eta=0$. \par
It should be noted that the occurrence of a thermodynamic transition 
in the one-dimensional model (\ref{eq:eHamiltonian})
does not imply a violation of well-known theorems:  
van-Hove's theorem \cite{vanHove} does not apply, since the 
Hamiltonian includes an on-site term; furthermore, since there is
no symmetry-breaking (or domain-wall-like solitons) involved,
the standard Landau argument against phase transitions 
in one dimension is also inapplicable.\par
Numerical results for a type-II model ($\rho=1,
\alpha=0.35 \AA^-1, \alpha/a=0.078$), obtained by 
solving the TI equation (\ref{eq:transferoperator}) for various
values of $T$ and $h$, are summarized in Fig. 1 \cite{Tc}.
Below $T_{c}$, the difference between the two lowest eigenvalues,
which determines $\xi_{||}$ as well as the singular part of the free
energy, is found empirically to satisfy the scaling equation
\begin{equation}
\label{Phi}
\epsilon_{1}-\epsilon_{0} = -f_{sing} =
D |t|\ \Phi\left(\frac{h}{|t|^{3/2}}\right)  \>,
\end{equation}
with the limiting forms $\Phi(x)=\Phi_{0}+\Phi_{1}x+\Phi_{2}x^{2}
+... (x\ll 1)$ and $\Phi(x) \propto x^{2/3} (x\gg 1)$;
it follows that $\nu_{||}=1$, $\gamma =2 $ and $-\beta=1/2$.
The values of the first two critical exponents are in very good
agreement with those  obtained directly from our zero-field results 
for the longitudinal correlation length and the susceptibility, 
respectively(cf. Fig. 2). The order parameter and, to a lesser extent, 
the transverse correlation length (also shown in Fig. 2) reveal 
significant deviations from pure power-law behavior 
- presumably due to strong transients (cf below); results are
however roughly consistent with $\nu_{\perp}=1/2$.\par
It is possible to follow  the transition from type-I to type-II 
behavior by continuously varying the stacking parameter 
$\alpha$, at constant $\rho=1$. We have done this using numerically
evaluated semiclassical (Bohr-Sommerfeld) eigenvalues of 
Eq. \ref{anhPS}. The results, shown
in Fig. 3, indicate that a crossover takes place at 
 $\alpha/a \approx  0.5$. Smaller values of  $\alpha/a$ correspond to
a longer -yet finite- range of the entropic barrier, compared
with that of the Morse potential; one obtains the type-II
exponent,  $\nu_{||}=1$. As  $\alpha/a$ increases and the range
of the entropic barrier becomes shorter than that of the 
Morse potential, the entropic barrier becomes irrelevant to
critical behavior; one obtains the type-I exponent, $\nu_{||}=2$.  \par
Equation (\ref{Phi}) states that the singular part of the
zero-field free energy depends linearly on the temperature.
In other words, the value of the exponent $\nu_{||}=1$ implies a
first-order transition. For our parameter set, the corresponding
melting entropy is $\Delta S/k_{B}= [D/(k_{B}T_{c})]\Phi_{0}=.75$
(cf inset in Fig. 4).\par
At this point a comment on the fraction of bound pairs is in order;
it is this quantity which one measures for DNA, using UV absorbance,
rather than the order parameter. In the type-I model, the probability
of finding a given base pair at an equilibrium distance smaller than $b$
(equal to the fraction of bound pairs, with a proper choice of $b$),
is given in terms of the incomplete gamma function, i.e.
$P(y<b,T)
= 1-\gamma(2\delta-1,2\delta e^{-ab}) / \Gamma(2\delta-1)$,
and approaches zero continuously as $T\to T_{c}$, independently
of the choice of $b$.
This is not the case in type II behavior. It can be seen in Fig. 4 that
there is a step discontinuity in the fraction of bound pairs; the exact
magnitude of the step depends on the choice of $b$, but the
discontinuity appears to be an intrinsic property of the type-II ground
state wave function; the subtlety
lies in the fact that, although there is a long-tail which causes the
divergence of $\sigma$, a finite weight of $|\phi_{0}|^2$ originates in
finite displacements.  This is to be contrasted with type-I behavior,
where more and more weight is
shifted to infinity as the transition is approached. Therefore,
the experimental detection of the fraction of bound pairs in terms of
the function $P(y<b,T)$, provides accurate information about the
true order of the transition; the order parameter
itself diverges smoothly at the transition and might, due
to transients which mask the leading-order asymptotics,
be less suited for a detailed study, even if it were readily available
by experimental methods. This is a fortunate natural coincidence.\par
In summary, we have presented a complete scaling analysis of a 
simple model which has been developed in order to describe 
the melting of "homogeneous" DNA. In terms of biophysical 
applications, our results should be complemented by the analysis 
of Ref.\cite{CuleHwa} to account for the effects
of heterogeneity. We feel nonetheless, that there is a {\em direct}
gain from the analysis presented here:
the melting of a double-stranded chain - a fairly
general problem of biologically motivated statistical
physics - has been shown to be a true thermodynamic transition with
different types of critical behavior, depending on the
details of the interaction parameters. It would be interesting to
speculate on whether other varieties of thermal biomolecular
denaturation (e.g. protein unfolding) could be studied in terms of 
related models of low-dimensional phase transitions.
\begin{figure}
\psfig{figure=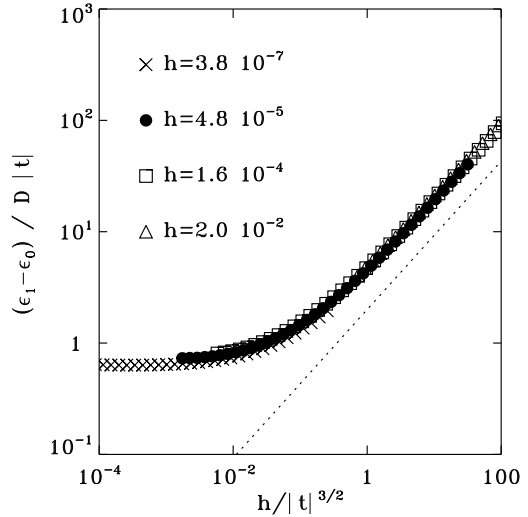,height=7.5truecm,width=7.5truecm}
\caption{Type II critical behavior: the dependence of $(\epsilon_{1}-
\epsilon_{0})/D|t|$ on
the scaling variable $h/|t|^{3/2}$ is shown for 4 different
values of $h$ and a range of temperatures; the dotted line marks the
slope 2/3. } 
\label{fig1}
\end{figure}
\begin{figure}
\psfig{figure=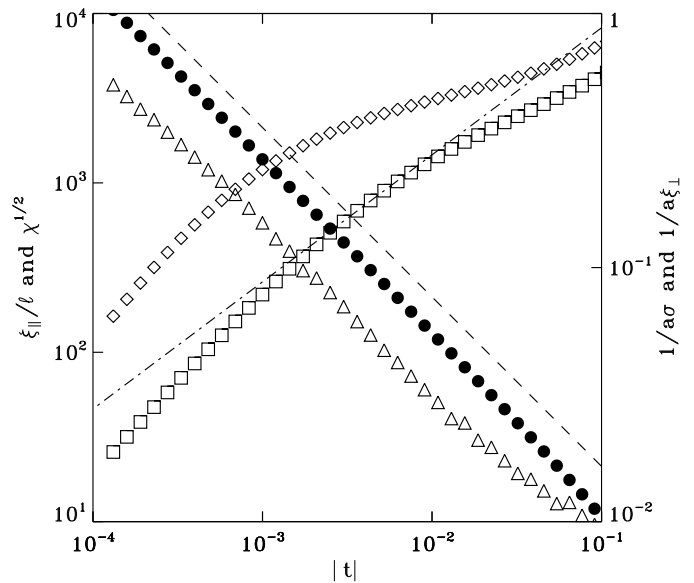,height=7.5truecm,width=7.5truecm}
\vskip .75truecm
\caption{Zero field results in the type-II case.
The longitudinal correlation length $\xi_{||}$ (circles, left
axis), the square root of the susceptibility $\chi$ (triangles,
left axis), the order parameter $\sigma$ (diamonds, right axis)
and its root-mean-square fluctuations $\xi_{\perp}$ (squares,
right axis) as a function of $|t|$. The dashed and dashed-dotted
lines correspond to $\nu_{||}=\gamma/2=1$ and $\nu_{\perp}=1/2$,
respectively.} \label{fig2}
\end{figure}
\begin{figure}
\psfig{figure=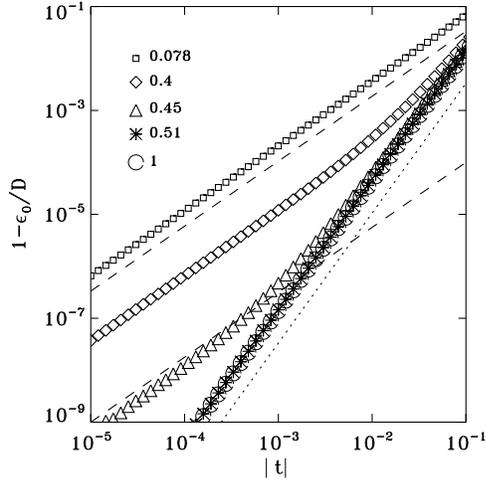,height=7truecm,width=7truecm}
\vskip .75truecm
\caption{The difference between the lowest eigenvalue 
of Eq. \ref{anhPS} and the bottom of continuum band, 
as a function of the reduced
temperature, calculated numerically for various values of 
the ratio $\alpha/a$, within
the Bohr-Sommerfeld quantization scheme. 
The asymptotic slope, which gives
the critical exponent $\nu_{||}$, 
changes from 1 to 2, as the stacking parameter 
varies from type-II ($\alpha/a<1/2$) to type-I ($\alpha/a>1/2$) values.
The dashed and dotted lines have slopes of 1 and 2, respectively.
} \label{fig3}
\end{figure}
\begin{figure}
\psfig{figure=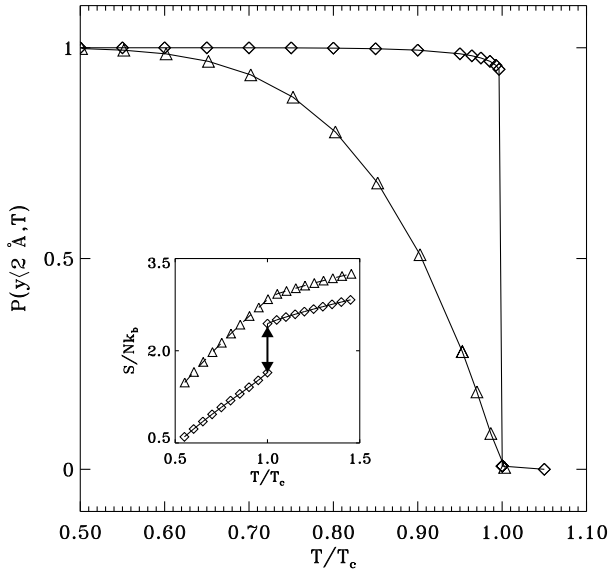,height=7truecm,width=7truecm}
\vskip .8truecm
\caption{ The fraction of bound base pairs
$P(y<2\AA,T)$ as a function of the $T/T_c$ for the type I
(triangles) and II (diamonds). Inset: the entropy $S(T)/Nk_B$)
(symbols as in main fig.); the length of the double arrow
represents the estimate of the melting entropy obtained from the
scaling Eq. (\ref{Phi}). The solid lines are "guides to the eye".}
\label{fig4}
\end{figure}


\begin{references}
\bibitem{details}
The details of the calculation will be reported elsewhere.
\bibitem{Tc}
The transition temperature
was estimated by exploiting the "topological" difference
occurring in a family of $\sigma(T,h)$ vs. $T$ plots for various
small fields: all the "low temperature" plots give a  finite limit as $h 
\to 0$; the "high temperature" plots give a divergent $\sigma(T,0)$. 
The procedure determines $T_{c}$ within a relative error
of $3\times10^{-5}$.
\end{references}
\end{document}